\documentclass[prl,aps,twocolumn,showpacs,superscriptaddress]{revtex4}
\usepackage{graphicx, amsmath, amsthm, subfigure}
\makeatletter
\begin{document}
\title{A Generic Two-band Model for Unconventional Superconductivity and Spin-Density-Wave Order in Electron and Hole Doped
Iron-Based  Superconductors}

\author{Qiang Han}
\affiliation{Department of Physics, Renmin University, Beijing, China}
\author{Yan Chen}
\affiliation{Department of Physics and Lab of Advanced Materials, Fudan
University, Shanghai, China}
\author{Z. D. Wang}
\email{zwang@hkucc.hku.hk}
 \affiliation{Department of Physics and
Center of Theoretical and Computational Physics, The University of
Hong Kong, Pokfulam Road, Hong Kong, China}

\date{\today}

\begin{abstract}

Based on experimental data on the newly synthesized iron-based
superconductors and the relevant band structure calculations, we
propose a minimal two-band BCS-type Hamiltonian with the interband
Hubbard interaction included. We illustrate that this two-band model
is able to capture the essential features of unconventional
superconductivity and spin density wave (SDW) ordering in this
family of materials. It is found that bound electron-hole pairs can
be condensed to reveal the SDW ordering for zero and very small
doping, while the superconducting ordering emerges at small finite
doping, whose pairing symmetry is qualitatively analyzed to be of
nodal $d$-wave. The derived analytical formulas not only give out a
nearly symmetric phase diagram for electron and hole doping, but
also is likely able to account for existing main experimental
results. Moreover, we also derive two important relations for a
general two-band model and elaborate how to apply them to determine
the band width ratio and the effective interband coupling strength
from experimental data.

\end{abstract}

\maketitle

Since the recent discovery of a new iron-based layered
superconductor~\cite{JACS}, intensive efforts have been focused on
the nature of superconductivity in this materials both
experimentally~\cite{NLWang1,HHWen1,HHWen2,Jin,HHWen3,NLWang2,NLWang3,HHWen4,ChenXH,ZXZhao,DLFeng}
and
theoretically~\cite{LDA0,LDA1,LDA2,Multi,LDA3,LDA4,DMFT,Phonon,XDai}.
Apart from the well-known copper oxide superconductors, this family
of materials exhibit higher critical temperatures, 26K in
LaO$_{0.9}$F$_{0.1}$FeAs~\cite{JACS}, 41K in
CeO$_{1-x}$F$_x$FeAs~\cite{NLWang2}, 43K in
SmO$_{1-x}$F$_x$FeAs~\cite{ChenXH}, and 52K in
PrO$_{0.89}$F$_{0.11}$FeAs~\cite{ZXZhao}, as well as 25K in hole
doped La$_{1-x}$Sr$_{x}$OFeAs~\cite{HHWen3}. Very recently, a number
of preliminary analyses have been made  to unveil the mystery of
superconducting nature, such as the multiband superconductivity,
unconventional pairing symmetry, electron-doping and hole-doping
effects, strong magnetic instability of the normal state.
Experiments from the specific heat measurements~\cite{HHWen1},
point-contact tunneling spectroscopy~\cite{HHWen2},  and infrared
reflectance spectroscopy~\cite{NLWang2} provided useful information.
For example, according to the point-contact tunneling spectroscopy
experiment by Shan {\em et al.}~\cite{HHWen2},  a remarkable
zero-bias conductance peak was observed at the (110) interface,
indicating the possible presence of nodal superconductivity. Ou {\em
et al.}~\cite{DLFeng} performed angle-integrated photoemission
spectroscopy measurements and their data provided certain support
for the existence of SDW ordering and an indication of
unconventional superconductivity. In the theoretical aspect, the
nature of unconventional superconductivity and the pairing mechanism
have also been explored preliminarily by several groups based on the
density functional theory (DFT) and dynamic mean filed theory
(DMFT)~\cite{LDA2,Multi,LDA3,LDA4,DMFT}. It was pointed out that the
electron-phonon interaction in this system may be too weak to lead
such high critical temperatures~\cite{Phonon}.
The possibility  of spin triplet superconductivity was also
suggested~\cite{XDai,wenxg}.

In this paper, we propose a minimal two-band BCS-type Hamiltonian
with an effective interband Hubbard interaction term included to
model the system. The construction of our model Hamiltonian is based
on band structure calculation results and intuitive physical
pictures.
 Taking into account the main
features of fermi surface for the undoped material calculated from
the DFT and to capture the essential physics of the
superconductivity and magnetism in the present system, we adopt a
minimal version of the Fermi surface on a primary two-dimensional
square lattice in the Fe-Fe plane: one hole band around $\Gamma$ and
one electron band around $M$ points, both crossing the Fermi surface
in the undoped case.
Based on rational physical considerations, we introduce an effective
interband antiferromagnetic interaction and elucidate that the
effective intraband antiferromagnetic coupling could induce the
superconducting pairing with a $d$-wave symmetry.
Our main findings are: (i) the normal state has an SDW order in the
undoped case, while upon the charge carrier doping the SDW order
drops rapidly and the superconducting order emerges; in particular,
analytical results for both SDW and superconducting transition
temperatures are explicitly presented, and their respective
relations to the SDW and superconducting gaps are elaborated in
details, with the former agreeing well with the existing experiment
and the latter being quite useful for the future experimental
verification; (ii) due to the two-band (electron and hole)
superconducting nature of the material, the transition temperature
as a function of the effective doping density shows a nearly
symmetric electron-hole doping dependence, accounting for the
experimental results; (iii) based on a reasonable analysis, the
two-band superconducting state is expected to possess a $d$-wave
pairing symmetry. Moreover, we address how to enhance
superconducting T$_c$ in this family of iron-based materials and how
to verify our two band model/theory without any fitting parameter.

We start from a minimal two-band  model, which captures the
essential physics of the multiband unconventional superconducting
state and SDW ordering,
\begin{eqnarray}
H&=&\sum_{k\sigma }\xi _{1k}c_{k\sigma }^{\dagger }c_{k\sigma }+
\sum_{k\sigma }\xi _{2k}d_{k\sigma }^{\dagger }d_{k\sigma }+
U_\mathrm{eff}\sum_{i,\sigma} n_{1i\sigma} n_{2i {\bar {\sigma}}}
\nonumber \\
&+&\sum_{kk^{\prime }}V^{11}_{kk^{\prime }}c_{k^{\prime }\uparrow
}^{\dagger }c_{-k^{\prime }\downarrow }^{\dagger }c_{-k\uparrow
}c_{k\downarrow} + \sum_{kk^{\prime }}V^{22}_{kk^{\prime
}}d_{k^{\prime }\uparrow }^{\dagger }d_{-k^{\prime }\downarrow
}^{\dagger }d_{-k\uparrow }d_{k\downarrow}
\nonumber \\
&+&\sum_{kk^{\prime }} \left( V^{12}_{kk^{\prime }}c_{k^{\prime
}\uparrow }^{\dagger }c_{-k^{\prime }\downarrow }^{\dagger
}d_{-k\uparrow }d_{k\downarrow } + h.c.\right) ,
 \label{ham1}
\end{eqnarray}
where the two band bare dispersions are respectively approximated as
$\xi _{1k}=-\frac{\hbar^2 k^2}{2m_1}+\epsilon^{(1)}_0 -\mu$ and $\xi
_{2k}=\frac{\hbar^2 ({\bf k-M})^2}{2m_2}-\epsilon^{(2)}_0 -\mu$
($\hbar=1$ hereafter) based on the band
calculations~\cite{LDA0,LDA1,LDA3,LDA4}, where $m_1=1/t_1$ and
$m_2=1/t_2$ are the effective masses of the hole and electron with
$t_{1,2}$ as the effective nearest neighbor hopping integrals in the
primary square lattice of sites (with each site as a unit cell
consisting of two Fe atoms), which is rotated by the angle of
$\pi/4$ and is enlarged by a factor of $\sqrt{2}\times \sqrt{2}$
with respect to the reduced Fe-Fe lattice, $\mu$ is the chemical
potential depending on filling, and $\epsilon^{(l)}_0$ stands for
the band offset, where $l=1,2$ represents respectively the nearly
filled valence band (hole band) around ${\bf \Gamma}=\left(0
,0\right) $ and the nearly empty conduction band (electron band)
around ${\bf M}=\left( \pi,\pi \right)$, as shown schematically in
Fig.\ 1.
\begin{figure}[htb]
\center
\includegraphics[width=8cm]{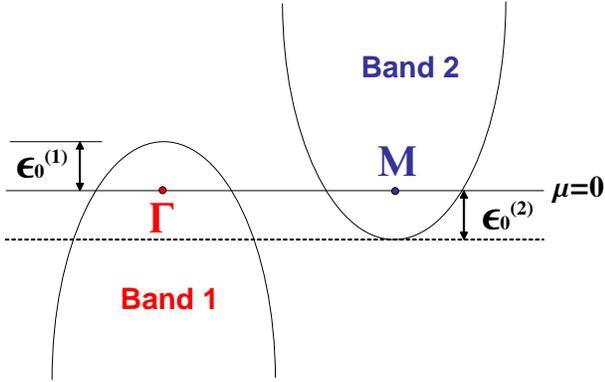}
\caption{Schematic plot of the bare band dispersions of the valence
(hole) and conduction (electron) bands for the undoped parent
compound ($\mu=0$). $\epsilon^{(1,2)}_0$ determines the initial
carrier density before doping.}
\end{figure}
$d_{k\sigma }$ and $c_{k\sigma }$ are the corresponding electron
annihilation operators of bands 1 and 2. For the hole(electron)
band, we have the density of states $\rho_{1,2}=1/(4\pi t_{1,2})$
with the band width $W_{h,e}=1/\rho_{1,2}$. $\epsilon^{(l)}_0$ is
set to give the carrier density at the undoped case, i.e. the hole
density of band-1 $n_h^0=2\rho_1\epsilon^{(1)}_0$, the electron
density of band-2 $n_e^0=2\rho_2\epsilon^{(2)}_0$, and  the
effective total electron number per site is $(2+n_e^0-n_h^0)$.
$V^{11}_{kk^{\prime }}$ and $V^{22}_{kk^{\prime }}$ are the
intraband pairing potentials for the two bands. $ V^{12}_{kk^{\prime
}}$ denotes the  interband pairing interaction. $U_\mathrm{eff}$
represents the effective interband Hubbard interaction term, which
will be elaborated in the next paragraph.

\begin{figure}[htb]
\center
\includegraphics[width=7.5cm,height=3cm]{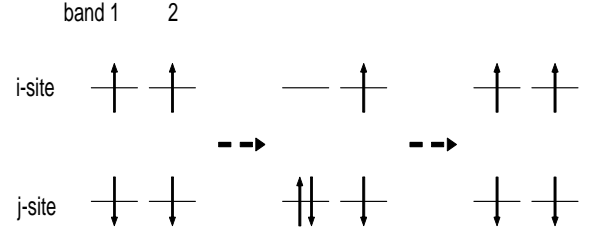}
\caption{Schematic representation of the origin of the
antiferromagnetic interaction.} \label{figure2}
\end{figure}

We now attempt to give an intuitive physical picture to understand
the origin of superconducting pairing and the SDW ordering in this
system. It is well known that the superexchange antiferromagnetic
coupling derived from the strong coupling  of the single-band
Hubbard model provides a driving force for the unconventional
$d$-wave superconducting pairing state in cuprate superconductors.
In the present case with the two bands, the $i$-th site Hamiltonian
is given by $H_i=U
n_{il\sigma}n_{il\bar{\sigma}}+U^{\prime}n_{i1\sigma}n_{i2\sigma\prime}-
J_\mathrm{H}$ {\boldmath$\sigma$}$_{i1}$ $\cdot$ {\boldmath$
{\sigma}$}$_{i2}$ with $n_i$ and {\boldmath${\sigma}$}$_i$ as the
$i$-site electron number and spin operators while $U$, $U^\prime$,
and $J_\mathrm{H}$ being respectively as the on-site intraband
Hubbard repulsion, interband one, and the Hund's coupling constant,
 which involves also the Hund's coupling and interband $U^\prime$ and thus
 introduces
more complications. To simply this complicated and subtle issue but
without loss of the key point, let us consider only the energy range
above the bottom of band 2 (i.e., $\mu -\epsilon_0^{(2)}$), below
which only the band 1 is available for the occupation by electrons
and thus the Hund's coupling plays no role.  The number of electrons
in this range at zero temperature is $ (1 +  \rho_h/\rho_e)n_e^0 N$
with $N$ as the number of sites. For the two empty neighboring
sites, each having two levels, there will be four levels available
for the occupation by four electrons in the considered energy range.
In the absence of hopping, due to the on-site Hubbard repulsion,
each site would have two electrons to lower the energy (see the left
panel of Fig.~2). In this case, there are six states, where the two
lowest energy states have two electrons with the parallel spins in
the two different bands (both up or down) due the Hund's coupling,
while the two relatively higher energy degenerate states have two
opposite spins in the two different bands, with the energy
difference $J_\mathrm{H}/2$ from the lowest energy level; the rest
two states have an even higher energy with opposite spins in either
of two bands as the intraband on-site $U$-repulsion is stronger than
the interband one ($U^\prime$)  (Here we refer to these six states
as being in the lower Hubbard band). When the hopping is included,
more very high energy states, with each site having at least three
electrons, are involved (we refer to these states as being in the
upper Hubbard band). If an electron hops from the bare ground state
via virtual processes to the upper Hubbard band and then back, which
involves an additional intraband $U$ and corresponds actually to the
antiferromagnetic superexchange-like process, lowers the total
ground state energy of the two sites. A typical hopping process is
schematically shown in Fig. 1. Analogous to the single-band case,
the present superexchange-like interaction could also lead to the
effective antiferromagnetic interaction between the two neighboring
sites and thus antiferromanetic fluctuations in the whole lattice
(as only a part of mobile electrons/sites are involved), which
introduce the effective superconducting pairing interaction having
most likely a nodal d-wave symmetry, as in cuprates. Also for
simplicity but with out loss of main physics, here we wish to argue
that only the interband term is retained~\cite{Rice} in the
effective Hamiltonian (1) after the virtual processes because the
intraband $U$ may be significantly greater than the interband
$U^{\prime}$.~\cite{note1} Since the interband repulsion term lifts
the energy $J_\mathrm{H}/2$ above the bare ground state energy
level, $U_\mathrm{eff}$ is $\sim J_\mathrm{H}/2$. Below we elaborate
first that this $U_\mathrm{eff}$-term can indeed lead to an SDW
order at zero or very small electron/hole doping, where the
superconducting paring order is suppressed.

The SDW transition temperature $T_\mathrm{SDW}$  can be calculated
from the following equation~\cite{Rice}
\begin{equation}
1=U_\mathrm{eff}\chi_0^{12}({\bf Q}),
\end{equation}
where $\chi_0^{12}$ is the interband spin susceptibility
\begin{equation}
\chi_0^{12}({\bf Q})=\sum_{{k}} \frac{f(\xi_{1{k}})-f(\xi_{2{
k+Q}})}{\xi_{1{k}}-\xi_{2{k+Q}}},\label{susc}
\end{equation}
with $f(\varepsilon)$ as the Fermi-Dirac function $
f(\varepsilon)=1/(1+e^{\varepsilon/T_c})$. To obtain a simple
analytic formula of $T_\mathrm{SDW}$, here we set $m_1=m_2$ and
$\epsilon^{(1)}_0=\epsilon^{(2)}_0=\epsilon_0$, where the perfect
nesting with vector ${\bf Q}=(\pi,\pi)$ between the two bands occurs
at the undoped case($\mu =0$), which leads to the SDW instability.
Integrating the RHS of Eq.~(\ref{susc}), we have an equation for
$T_\mathrm{SDW}$
 \begin{equation}
\frac{T_\mathrm{SDW}}{W}\approx\frac{2e^\gamma}{\pi}
\sqrt{\frac{\epsilon_0}{W}\left(1-\frac{\epsilon_0}{W}\right)}
e^{-\left(\frac{U_\mathrm{eff}}{W}\right)^{-1}}
e^{-1.71\left(\frac{W }{8T_\mathrm{SDW}}x \right)^2} , \label{tneel}
\end{equation}
where $\gamma\approx 0.577$ is the Euler constant, $x$ is the
effective small doping, and the condition $\epsilon_0\gg
T_\mathrm{SDW}$ has been used. When the electron or hole doping is
zero, i.e. $x=0$, we have the largest $T_\mathrm{SDW}$. Remarkably,
$T_\mathrm{SDW}$ drops drastically whenever more electrons or holes
are doped, as seen clearly in Eq.{(\ref{tneel}) and Fig.~3. If
$m_1\neq m_2$, $T_\mathrm{SDW}$ is expected to be lowered.

Below $T_\mathrm{SDW}$, the SDW ordering emerges, whose order
parameter may be defined as
\begin{equation}
\Delta_\mathrm{SDW}=\frac{U_\mathrm{eff}}{2}\sum_{k\sigma}\left\langle
c_{k\sigma}d^{\dagger}_{k+Q\bar{\sigma}}\right\rangle,
\label{deltasdw}
\end{equation}
where $\langle \cdots \rangle$ denotes thermodynamic average.
$\Delta_\mathrm{SDW}$ satisfies the following equation
\begin{equation}
1=U_\mathrm{eff}\sum_{k}
\frac{f(\eta_{2k}+\Omega_k)-f(\eta_{2k}-\Omega_k)}{2\Omega_k},
\end{equation}
where $\Omega_k=\sqrt{\eta_{1k}^2+\Delta_\mathrm{SDW}^2}$,
$\eta_{1k}=(\xi_{1k}-\xi_{2k+Q})/2$ and
$\eta_{2k}=(\xi_{1k}+\xi_{2k+Q})/2$. Here the effective repulsive
interaction ($U_\mathrm{eff}$) between interband electrons may also
be viewed as an attractive pairing interaction between electron and
hole. Physically, it is noted that the pairing leads to form a
"condensate" of bound electron-hole pairs in the triplet state or
"excitons"~\cite{Rice2}, which exhibits the SDW ordering.  The
condensate of electron-hole pairs is actually an counterpart of
Cooper electron-electron pairs. In this sense, it is straightforward
to obtain the famous relation $2\Delta_\mathrm{SDW}(0)\approx 3.5
T_\mathrm{SDW}$, as in the case of the conventional weak-coupling
superconductivity. Here we pinpoint out that this relation can be
quantitatively verified by independent experimental methods, e.g.,
the optical conductivity spectra~\cite{NLWang3} and resistivity or
specific heat measurements. As seen from Ref.\cite{NLWang3}, it may
be estimated that $2\Delta_\mathrm{SDW}(8\mathrm{K})\approx 350
\mathrm{cm}^{-1}=504 \mathrm{K}$ and $T_\mathrm{SDW}\approx 150
\mathrm{K}$, leading to
$2\Delta_\mathrm{SDW}(8\mathrm{K})/T_\mathrm{SDW}\approx 3.4$, which
agrees well with the present theory. It is worthwhile noting that,
since the effective hopping integrals $t_{1,2}$ in the present model
correspond to those between nearest neighboring sites (unit cells)
in the primary lattice, the above SDW ordering pattern is
stripe-like antiferromagnetic on the reduced Fe-Fe square lattice,
namely, the spin ordering pattern is ferromagnetic in each stripe
along the x'(or y')-direction of the reduced Fe-Fe lattice, while it
is antiferromagnetic between stripes in the y'(or x')-direction.
This kind of SDW ordering was likely observed in a very recent
neutron scattering experiment~\cite{Dai}. Remarkably, we estimate
from Eq.~(\ref{deltasdw}) that the antiferromagnetic moment per
Fe-atom (in unit of Bohr magneton $\mu_\mathrm{B}$):
$m_\mathrm{AF}\approx(4/2)\times
\overline{m_i}=2\sum_{k\sigma}\langle
c_{k\sigma}d^\dagger_{k+Q\bar{\sigma}}\rangle=4\Delta_\mathrm{SDW}/U_\mathrm{eff}=2\times3.52
\,(T_c/W)/(U_\mathrm{eff}/W)\approx 0.31$, where  $m_i=(-1)^i
\sum_\sigma\langle c_{i\sigma}d^\dagger_{i\bar{\sigma}}\rangle$ is
the $i$-th site moment~\cite{note2}. This estimation is also in
agreement with the data reported in Ref~\cite{Dai}. In addition, we
may have one more expectation from Eq.(\ref{tneel}) that $T_{SDW}$
is normally decreased with the increase of pressure because the
effective band width $W$ (or the effective hopping integral $t$)
that dominates $T_{SDW}$ in the exponential term is increased, as
seen in Ref.~\cite{chu}.

At this stage, we turn to address the superconducting ordering. In
view of the fact that the SDW order drops very sharply to zero at a
very small critical doping, it is reasonable and convenient to
ignore the effect of SDW order on the superconducting ordering above
the critical doping level.  In the following calculations, the
pairing potentials involving two bands are expressed as:
$V^{11}_{kk^{\prime }}=J_{hh}\gamma _{k}\gamma _{k^{\prime }}$ ,
$V^{22}_{kk^{\prime }}=J_{ee}\gamma _{k}\gamma _{k^{\prime }}$ and $
V^{12,21}_{kk^{\prime }}=J_{he,eh}\gamma_{k}\gamma _{k^{\prime }}$,
where $J_{hh}$, $J_{ee}$, and $J_{he,eh}$ are the corresponding
coupling constants, $\gamma _{k} = \cos (L \theta_k)$ with $L=2,1,0$
denoting respectively the nodal d-wave, nodal p-wave, and isotropic
s-wave pairing functions, and $\theta_k$ being the angle between the
vector $\bf{k}$ and the x-axis. Taking the BCS mean field
approximation, the quasiparticle eigenspectrum of the \textit{l}-th
band is then given by  $ E_{lk}=\sqrt{\xi_{lk}^{2}+|\Delta
_{l}|^{2}\gamma _{k}^{2}}$, where $|\Delta _{l}|$ is the gap
amplitude of the {\it l}-th band($l=h,e$). They are determined from
the following coupled gap equations~\cite{Suhl}:
\begin{eqnarray}
\Delta _{h} = \sum_{k}\gamma _{k}(J_{hh}\langle c_{-k\downarrow
}c_{k\uparrow }\rangle +J_{he}\langle d_{-k\downarrow }d_{k\uparrow
}\rangle), \nonumber \\
 \Delta _{e} = \sum_{k}\gamma _{k}(
J_{ee}\langle d_{-k\downarrow }d_{k\uparrow }\rangle + J_{eh}\langle
c_{-k\downarrow }c_{k\uparrow
}\rangle ) .  \label{gaph}
\end{eqnarray}
Then the self-consistent gap equations for $\Delta_h$ and $\Delta_e$
read
\begin{equation}
\left(
\begin{array}{cc}
J_{hh}K_1 & J_{he}K_2 \\
J_{eh}K_1 & J_{ee}K_2
\end{array}
\right)%
\left(\begin{array}{c}
\Delta_h \\
\Delta_e %
\end{array}%
\right)=%
\left(\begin{array}{c}
\Delta_h \\
\Delta_e %
\end{array}%
\right),%
\label{gapeq}
\end{equation}
where $ K_{1,2} = \sum_{k}
\gamma_{k}^2\tanh\left(E_{1,2k}/2T\right)/E_{1,2k}$ satisfy the
following equation

\begin{equation}
\det\left(%
\begin{array}{cc}
J_{hh}K_1-1 & J_{he}K_2 \\
J_{eh}K_1 & J_{ee}K_2-1
\end{array}
\right)=0.%
\label{Keq}
\end{equation}

The superconducting transition temperature $T_c$ is actually
determined from  Eq.~(\ref{Keq}) with $\Delta_l \rightarrow 0$,
which depends on the pairing interaction strengths and the normal
state dispersions of the two bands.

\begin{figure}[htb]
\begin{center}
\includegraphics[width=8cm]{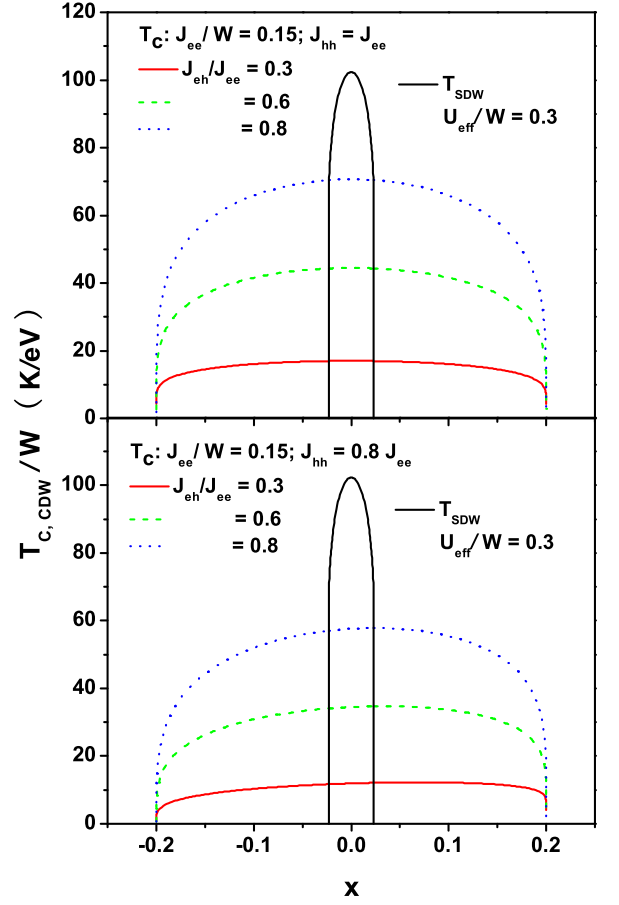}
\caption{(Color online) Superconducting critical temperature $T_c$
and the SDW transition temperature $T_\mathrm{SDW}$ as functions of
the effective doping $x$. The $T_\mathrm{SDW}$ curve is calculated
from Eq.\ (\ref{tneel}) with the parameters $U_\mathrm{eff}/W=0.3$
and $\epsilon_0/W=0.05$;  the $T_c$ curves are plotted by simply
substituting the relevant reduced parameters into Eq.\ (\ref{tc}),
where $J_{ee}/W=0.15$, and $|J_{eh}|/J_{ee}= 0.3, 0.6, 0.8$,
respectively for $J_{hh}=J_{ee}$ and $J_{hh}=0.8 J_{ee}$ .}
\end{center}
\end{figure}

Let us first address a more general case: $J_{ee},J_{hh} \geq 0$,
and $J_{ee}J_{hh}\neq J_{eh}J_{he}>0$ (noting that
$J_{eh}=J^{*}_{he}$). By the introduction of dimensionless
couplings:
$\tilde{J}_{hh,ee}=J_{hh,ee}/W_{h,e},
\tilde{J}_{eh,he}=J_{eh,he}/W_{h,e}$, and
$\widetilde{JJ}=\tilde{J}_{eh}\tilde{J}_{he} -
\tilde{J}_{ee}\tilde{J}_{hh}$,
 the $T_c$-formula is derived as
\begin{equation}
\frac{T_c}{\sqrt{W_e W_h}}=e^{C_{T_c}}\left[{n_e n_h
(2-n_e)(2-n_h)}\right]^{\frac{1}{4}}
e^{-\frac{1}{\lambda_\mathrm{red}}}, \label{tc}
\end{equation}
with $C_{T_c}=\ln (e^\gamma/\pi)\approx -0.568$ (for $L=2,1,0$
cases) and $\lambda_\mathrm{red}$ as the reduced pairing strength
being given by
\begin{eqnarray}
\lambda_\mathrm{red}^{-1}&=& \left\{
\left[\left(\frac{1}{4}\widetilde{JJ}\ln\frac{n_e(2-n_e)W_e^2}{n_h(2-n_h)W_h^2}
+ \frac{\tilde{J}_{hh}-\tilde{J}_{ee}}{2} \right)^2+ \right. \right.
\nonumber \\
&& \left. \left. \tilde{J}_{eh}\tilde{J}_{he}\right]^{\frac{1}{2}}
-\frac{1}{2}(\tilde{J}_{ee}+\tilde{J}_{hh})\right\} / \widetilde{JJ}
\nonumber,
\end{eqnarray}
where $n_e=n_e^0+[W_h/(W_e+W_h)]x$ and $n_h=n_h^0-[W_e/(W_e+W_h)]x$,
and the condition $\epsilon^{(1,2)}_0\gg T_c$ has been used. For a
special case: $\widetilde{JJ}=0$, by taking the limit of
$\widetilde{JJ}\rightarrow 0$ in the above equation, the
$T_c$-formula is reduced to
\begin{eqnarray}
\frac{T_c}{\sqrt{W_e
W_h}}&=&\frac{e^\gamma}{\pi}\left(\sqrt{\frac{W_e}{W_h}}\right)^{\frac{\tilde{J}_{ee}-\tilde{J}_{hh}}{\tilde{J}_{ee}
+ \tilde{J}_{hh}}}\left[\sqrt{{n_e
(2-n_e)}}\right]^{\frac{\tilde{J}_{ee}}{\tilde{J}_{ee} +
\tilde{J}_{hh}}} \nonumber \\
&\times&\left[\sqrt{{n_h
(2-n_h)}}\right]^{\frac{\tilde{J}_{hh}}{\tilde{J}_{ee} +
\tilde{J}_{hh}}}
e^{-\frac{1}{\tilde{J}_{ee}+\tilde{J}_{hh}}},\label{tcmax}
\end{eqnarray}
In this special case, more intriguingly, it is found from
Eq.~(\ref{gapeq}) that the ratio between two gaps $r(T)
=\Delta_{h}/\Delta_{e}=J_{he}/J_{ee}=J_{hh}/J_{eh}$, being
independent of temperature and other variables. Therefore, if the
ratio value $|r|$ is experimentally found to be independent of
temperature and other variables, it is implied that the system is
just in this special case. Note that this result is valid for any
two-band superconductivity model described by Eq.~(\ref{gaph}).

For a general case, i.e., $\widetilde{JJ}\neq 0$, the phase diagram
of $T_c$-$x$ calculated from above formula is plotted in Fig.~3,
where we choose the parameters as $W_e=W_h=W$ (for simplicity but
without loss of generality), $J_{ee}/W=0.15$, and
$\epsilon^{(1,2)}_0=\epsilon_0=0.05W$, $U_\mathrm{eff}/W=0.3$ (since
$J_\mathrm{H} \approx 0.9$ eV, i.e., $U_\mathrm{eff}\sim 0.45$ eV,
such choice leads to the effective band width $W\sim 1.5$ eV, which
is not unreasonable in the present system). From Fig.~3, we note
that the superconducting order emerges from a very small effective
doping and then decays rather slowly to zero at $x = x_{(e,h)}^c=\pm
2n^0_{(h,e)}$. When $J_{ee}=J_{hh}$ and $n^0_h=n^0_e$, the $T_c$
{\it vs.} $x$ dependence is symmetric for the effective electron and
hole doping ; while the symmetric feature changes slightly if
$J_{ee}$ (or $n^0_h$) is different from $J_{hh}$ (or $n^0_e$ ) not
too much (as estimated from the band calculations), in agreement
with the experimental result~\cite{HHWen3}. In particular, if one
intraband pairing strength (or the interband coupling) with the
largest value is fixed, both the interband coupling (regardless of
its sign) and other intraband pairing strength(s) enhance $T_c$
significantly, reaching the maximum as $|J_{eh}|\rightarrow J_{hh}
\rightarrow J_{ee}$, whose value is explicitly given by Eq.
(\ref{tcmax}). This feature may be helpful for searching even higher
$T_c$ superconductors.
Based on our results, another possible way to increase $T_c$ in this
family of superconductors is to increase the effective band width
$W$ (or the effective hopping integral $t$ ) since we have
approximately $T_c \propto W e^{-\alpha /W}$ with $\alpha$ being a
$W$-independent coefficient by noting that $J_{ee,hh}/W \propto W$.
In this sense, a higher pressure may enhance $T_c$, in contrast to
that of $T_{SDW}$. This expectation was seen in a very recent high
pressure experiment that showed clearly the enhancement of $T_c$
with the shrinkage of the lattice~\cite{pressure}.

\begin{figure}[htb]
\center
\includegraphics[width=8cm]{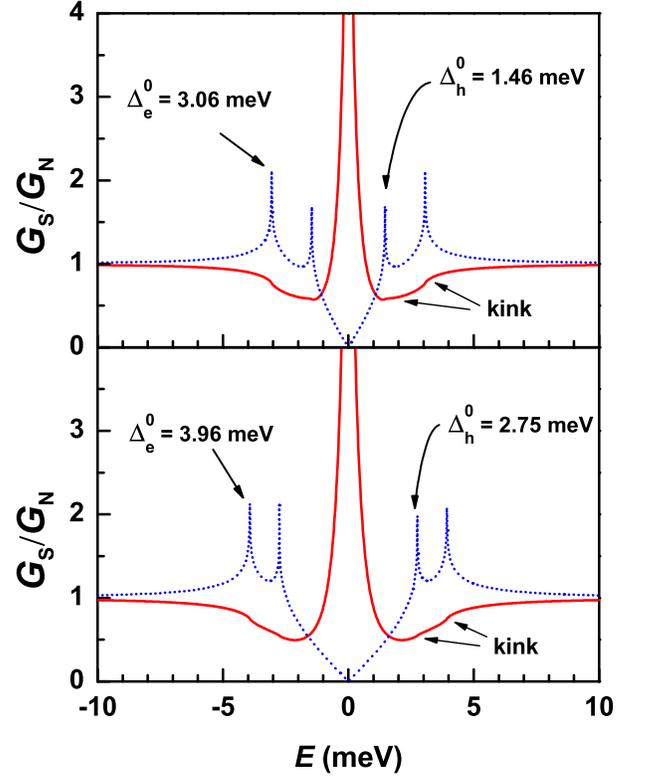}
\caption{(Color online) Normalized tunneling conductance as a
function of bias voltage at zero temperature for an electron doping
at $x=0.1$. The parameters are $W_h=W_e=1.5$ eV, $J_{ee}/W=0.15$,
$|J_{eh}|/J_{ee}= 0.3$ and $J_{hh}/J_{ee}=0.8$ for the lower panel,
which gives rise to the self-consistent $\Delta_e^0= 3.93$ meV and
$\Delta_h^0=2.75$ meV; for the upper panel $J_{hh}/J_{ee}=0.5$,
which leads to $\Delta_e^0\doteq3.06$ meV and $\Delta_h^0\doteq1.46$
meV.} ,
\end{figure}
  The normalized tunneling conductances along the directions (110)
and (001) of the primary lattice versus the bias voltage are
calculated self-consistently from the gap equations~(\ref{gapeq})
with the nodal d-wave ($\cos (2 \theta_k)$) pairing at zero
temperature and electron doping at $x=0.1$, as shown in Fig.~4 for a
set of parameters used in plotting Fig~3 (except for the upper panel
where $J_{hh}=0.5J_{ee}$). A sharp zero-bias-conductance-peak (ZBCP)
along (110) and two coherence peaks corresponding to the two gaps (a
larger $\Delta^0_{e}$ and a smaller $\Delta^0_{h}$) are clearly
seen, as expected. In addition, the two weak kinks in the ZBCP curve
can also be seen, which correspond to the two gaps as well. The
present ZBCP results are in agreement with the experimental
observation reported in Ref.~\cite{HHWen2}

 Finally, we wish to address how to verify unambiguously the present
 two band model/theory experimentally, without any fitting parameter.
 From the definitions of $K_{1,2}$ below
 Eq.~(\ref{gaph}), we have
\begin{eqnarray}
K_{1,2}(T_c)=\frac{
\ln\sqrt{n_{h,e}(2-n_{h,e})}-\ln\left(T_c/W_{h,e}\right)+C_{T_c}}{W_{h,e}},
\nonumber
\\
K_{1,2}(0)=\frac{\ln\sqrt{n_{h,e}(2-n_{h,e})}-
\ln(|\Delta^{0}_{h,e}|/W_{h,e})+C_{0}}{W_{h,e}}, \nonumber
\end{eqnarray}
where $C_0=(\ln2-1/2)\approx 0.193$ for the nodal d-wave and p-wave
cases (while it is zero for the gapped s-wave and p-wave cases). On
the other hand, from Eq.~(\ref{gapeq}), we have
$W_{h,e}K_{1,2}(T)=[-\tilde{J}_{ee,hh}+r^{-1,1}(T)\tilde{J}_{he,eh}]
/\widetilde{JJ}$ for any temperature $T \leq T_c$. Combining them,
we can find an important relation:
\begin{equation}
\frac{|-\ln (|\Delta^0_{h}|/T_c)+C_0-C_{T_c}|}{|-\ln
(|\Delta^0_{e}|/T_c)+C_0-C_{T_c}|}\times |r_0 r_c|=\frac{W_h}{W_e},
\label {rela1}
\end{equation}
where $r_{0,c}=r(0), r(T_c)$ are respectively the above introduced
gap ratio $r(T)$ below Eq.~(\ref{tcmax}) at zero and transition
temperatures. Since the RHS of Eq.~(\ref{rela1}) depends only on the
ratio of the two band widths, this relation can be verified by
experimentally measured data for $|\Delta^0_{e,h}|$ and $|r_{0, c}|$
at various doping levels and then be used to determine the ratio of
$W_h/W_e$. Similarly, we can obtain another useful relation:
\begin{equation}
\frac{\ln |r_0|}{[1+(W_h/W_e)|r_0 r_c|^{-1}]
(|r_c|-|r_0|)}=\frac{|\tilde{J}_{eh}|}{\widetilde{JJ}}. \label
{rela2}
\end{equation}
Once $(W_h/W_e)$ is determined from Eq.~(\ref{rela1}), this relation
can not only  be checked by experimental data of $r_{0,c}$ but also
be used to determine the effective interband coupling strength
$|\tilde{J}_{eh}|/\widetilde{JJ}$. It is also interesting to see
from Eq.~(\ref{rela2}) that (i) as long as $|r_0|\approx |r_c|$ is
observed experimentally, the system is approximately in the
mentioned special case; (ii) if $|r_0|\approx 1$ is seen
experimentally, except the case (i), the interband coupling is
negligible; (iii) if two bands superconduct independently with two
different $T_c$'s, the system $T_c$ is determined by the band having
a higher one, and the LHS of Eq.~(\ref{rela2}) approaches to zero
because $r_c$ is either divergent or zero at the lower $T_c$. It is
notable that the above results are also valid for the two-band
s-wave and gapped p-wave superconductors  similar to the present
system, except for a different constant term
$(C_0-C_{T_c})_{s,p}\approx 0.568$ .

In summary, we have proposed for the first time a minimal two-band
(hole and electron) model, with the interband Hubbard interaction
being also included. It has been shown that this simplified model is
able to capture the essential physics of unconventional
superconductivity and spin density wave ordering in the addressed
new family of materials. The present theory not only accounts
 for very recent experimental results, including the emergence
 of hole-doped superconducting states and
the neutron scattering analysis on the SDW ordering, but also
demonstrates/elaborates a key role of the interband pairing coupling
played in the significant enhancement of $T_c$ (in comparison with
that of each band) for a generic two-band system, regardless of the
pairing symmetry.


\acknowledgments We gratefully acknowledge helpful discussions with
Profs.
Y. P. Wang, N. L. Wang, D. L. Feng, J. X. Li, and J. Shi. 
This work was supported by the NSFC grand under Grants Nos. 10674179
and 10429401, the GRF grant of Hong Kong,  the Croucher Senior
Research Fellowship, and the Universitas 21 Fellowship. ZDW would
like to thank Peking University and the Institute of Physics (CAS)
for their hospitality during the visiting period when this work was
completed.


\begin{thebibliography}{20}
\bibitem{note1} If $(U-U^{\prime})\approx U_\mathrm{eff}$, the
effective on-site intraband repulsive terms
$((U-U^{\prime}+J_H/4)\sum_{i,l, \sigma} n_{li\sigma} n_{li {\bar
{\sigma}}}$ may need to be considered, which will be addressed
elsewhere later. On the other hand, we wish to emphsize that the
conventional s-wave superconducting pairing is unfavored due to the
above on-site intraband repulsive terms, as in cuprates.
\bibitem{note2} Considering that the almost degenerated two hole
bands and two electron bands connected to two Fe atoms per site and
the $m_i$
 counts merely the contribution per site from one nesting channel between one hole
 and one electron bands, while there are four such independent channels,  we have approximately
 the factor (4/2) here.
\end{thebibliography}
\end{document}